\documentclass[aps,prb,twocolumn,showpacs,preprintnumbers,superscriptaddress,amsmath,amssymb]{revtex4-1}

\usepackage[utf8]{inputenc} % symbols for accents
\usepackage[english]{babel} % language used
\usepackage{graphicx}% Include figure files
\usepackage{amssymb,amsmath} % math tools
\usepackage[hidelinks]{hyperref} % hyperlink to references
\usepackage{epstopdf}
\usepackage{nicefrac}
\usepackage{dcolumn}% Align table columns on decimal point
\usepackage{bm}% bold math
\usepackage{epsfig}
\usepackage{color}
\usepackage{siunitx} % SI units symbols
\usepackage{multirow}
\usepackage{physics}
\hypersetup{
	colorlinks,
	citecolor=blue,
	linkcolor=blue,
	urlcolor=blue,
}

\usepackage{tikz}
\usepackage{braket}
\newcommand*{\tikzstate}[2]{\tikz[baseline={([yshift=-.5ex]current bounding box.center)}]{\draw (0,0) pic{#1};\draw (.5,0) pic{#2};}}

\tikzset{
       lowup/.pic={
	  \begin{scope}[scale=.4]
	     \draw[white] (0,-.7) rectangle (1,.75);
	  \draw  (0,-.3) -- (1,-.3) ;
	  \draw  (0,.3) -- (1,.3) ;
	  \draw [ thick, -{stealth[]}, blue] (.5,-0.65) -- (.5,0.1) ;
       \end{scope}
       }
    }
\tikzset{
       highup/.pic={
	  \begin{scope}[scale=.4]
	     \draw[white] (0,-.7) rectangle (1,.75);
	  \draw  (0,-.3) -- (1,-.3) ;
	  \draw  (0,.3) -- (1,.3) ;
	  \draw [ thick, -{stealth[]}, blue] (.5,-0.05) -- (.5,0.7) ;
       \end{scope}
       }
    }
\tikzset{
       highdown/.pic={
	  \begin{scope}[scale=.4]
	     \draw[white] (0,-.7) rectangle (1,.75);
	  \draw  (0,-.3) -- (1,-.3) ;
	  \draw  (0,.3) -- (1,.3) ;
	  \draw [ thick, {stealth[]}-, blue] (.5,-0.1) -- (.5,0.65) ;
       \end{scope}
       }
    }
\tikzset{
       lowdown/.pic={
	  \begin{scope}[scale=.4]
	     \draw [white] (0,-.7) rectangle (1,.75);
	  \draw  (0,-.3) -- (1,-.3) ;
	  \draw  (0,.3) -- (1,.3) ;
	  \draw [ thick, {stealth[]}-, blue] (.5,-0.7) -- (.5,0.05) ;
       \end{scope}
       }
    }
\begin{document}

\title{Mobile orbitons in Ca$_2$CuO$_3$: crucial role of the Hund's exchange}

\author{R. Fumagalli}
\affiliation{Dipartimento di Fisica, Politecnico di Milano, Piazza Leonardo da Vinci 32, I-20133 Milano, Italy}

\author{J. Heverhagen}
\affiliation{Institute for Functional Matter and Quantum Technologies, University of Stuttgart, Pfaffenwaldring 57, D-70550 Stuttgart, Germany}

\author{D. Betto}
\altaffiliation[Present address: ]{Max Planck Institute for Solid State Physics, Heisenbergstra{\ss}e 1, D-70569 Stuttgart, Germany}
\affiliation{ESRF, The European Synchrotron, BP 220, F-38043, Grenoble Cedex, France}

\author{R. Arpaia}
\affiliation{Dipartimento di Fisica, Politecnico di Milano, Piazza Leonardo da Vinci 32, I-20133 Milano, Italy}
\affiliation{Quantum Device Physics Laboratory, Department of Microtechnology and Nanoscience, Chalmers University of Technology, SE-41296 G\"{o}teborg, Sweden}

\author{M. Rossi}
\altaffiliation[Present address: ]{Stanford Institute for Materials and Energy Sciences, SLAC National Accelerator Laboratory and Stanford University, 2575 Sand Hill Road, Menlo Park, CA-94025, USA}
\affiliation{Dipartimento di Fisica, Politecnico di Milano, Piazza Leonardo da Vinci 32, I-20133 Milano, Italy}

\author{D. Di Castro}
\affiliation{Dipartimento di Ingegneria Civile e Ingegneria Informatica, Universita` di Roma Tor Vergata, Via del Politecnico 1, I-00133 Roma, Italy}
\affiliation{CNR-SPIN, DICII, Università di Roma Tor Vergata, Via del Politecnico 1, Roma, I-00133, Italy}

\author{N.B. Brookes}
\affiliation{ESRF, The European Synchrotron, BP 220, F-38043, Grenoble Cedex, France}

\author{M. Moretti Sala}
\affiliation{Dipartimento di Fisica, Politecnico di Milano, Piazza Leonardo da Vinci 32, I-20133 Milano, Italy}

\author{M. Daghofer}
\affiliation{Institute for Functional Matter and Quantum Technologies, University of Stuttgart, Pfaffenwaldring 57, D-70550 Stuttgart, Germany}
\affiliation{Center for Integrated Quantum Science and Technology, University of Stuttgart, Pfaffenwaldring 57, 70550 Stuttgart, Germany}

\author{L. Braicovich}
\affiliation{Dipartimento di Fisica, Politecnico di Milano, Piazza Leonardo da Vinci 32, I-20133 Milano, Italy}
\affiliation{ESRF, The European Synchrotron, BP 220, F-38043, Grenoble Cedex, France}

\author{K. Wohlfeld}
\email{krzysztof.wohlfeld@fuw.edu.pl}
\affiliation{Institute of Theoretical Physics, Faculty of Physics, University of Warsaw, Pasteura 5, PL-02093 Warsaw, Poland}

\author{G. Ghiringhelli}
\email{giacomo.ghiringhelli@polimi.it}
\affiliation{Dipartimento di Fisica, Politecnico di Milano, Piazza Leonardo da Vinci 32, I-20133 Milano, Italy}
\affiliation{CNR-SPIN, Dipartimento di Fisica, Politecnico di Milano, 20133 Milano, Italy}

\date{\today}

\begin{abstract}
We investigate the Cu $L_3$ edge resonant inelastic x-ray scattering (RIXS)
spectra of a quasi-1D antiferromagnet Ca$_2$CuO$_3$. In addition to the magnetic
excitations, which are well-described by the two-spinon continuum, we observe
two dispersive orbital excitations, the $3d_{xy}$ and the $3d_{yz}$ orbitons.
We carry out a quantitative comparison of the RIXS spectra, obtained with two distinct incident polarizations, with a theoretical model. We show that any realistic spin-orbital model needs to include a finite, but realistic, Hund’s exchange $J_H \approx 0.5$ eV. Its main effect is an increase in orbiton velocities, so that their theoretically calculated values match those observed experimentally.
Even though Hund's exchange also mediates some interaction between spinon and orbiton, the picture of spin-orbit separation remains intact and describes orbiton motion in this compound.
\end{abstract}

\maketitle

\section{Introduction}
\label{sec:introduction}
The importance of the Hund's exchange and the multi-orbital character of the states lying close to the Fermi level go hand in hand in the correlated electron systems~\cite{Kugel1982}.
In addition to `Hund's metallicity' in itinerant systems~\cite{Georges2013},
there are Mott-insulating transition-metal compounds with almost degenerate orbitals, which can show spin and orbitally ordered ground states~\cite{Kugel1982, Tokura2000}.
Interestingly, the spin and orbital order in these compounds often follows a special kind of `complementarity rule', typically known as the Goodenough-Kanamori rule~\cite{Kanamori1959, Goodenough1963}: the bond with a dominant alternating orbital (ferro-orbital) correlation shows ferromagnetic (antiferromagnetic) correlation, respectively.

\begin{figure}[htbp]
	\includegraphics[width=1\columnwidth]{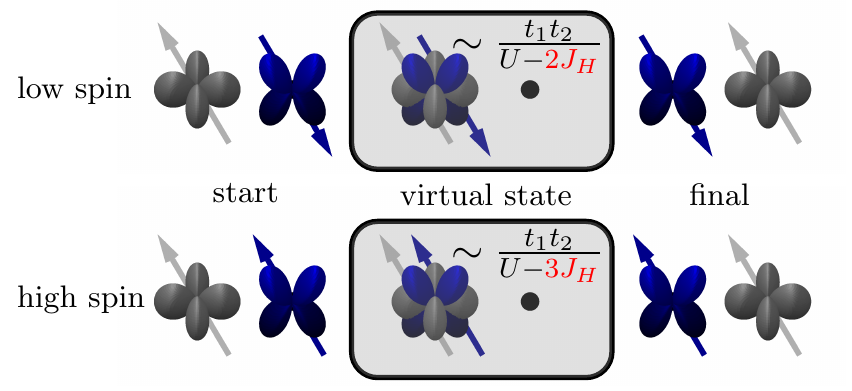}
	\caption{\label{fig:O_SO_mechanism}
		A cartoon picture of the two possible spin-orbital exchange processes along a bond with two distinct orbitals occupying
		the nearest neighbor sites. Whereas in the limit of vanishing Hund's exchange $J_H$ the amplitude of such processes is solely $\propto t_1 t_2 / U$
		(with $t_1, t_2$ being the nearest neighbor hoppings between respective orbitals and $U$ the Coulomb repulsion element on the same orbital),
		in a `realistic' case it also depends on $J_H/U$ and is larger (smaller) for the the parallel (antiparallel) spin alignment as presented on the bottom (top) panels, respectively.
		This not only explains the origin of one of the Goodenough-Kanamori rules (see Sec.~\ref{sec:introduction}) but also the
		dependence of the orbiton velocity on the spin correlations as well as its overall increase with the Hund's exchange, as discussed in detail in this paper.
	}
\end{figure}

Consequently and especially once the typically strong Jahn-Teller effect is included, various transition metal oxides or fluorides show strongly
anisotropic magnetic ordering~\cite{Khomskii2014} -- probably the most famous example is the cubic LaMnO$_3$ with its
$ab$ planes (chains along the $c$ axis) showing ferromagnetic (antiferromagnetic) order~\cite{Moussa1996, Oles2005}.

The Goodenough-Kanamori rules are intimately related to the Hund's exchange.
The crucial observation is that Hund's exchange is responsible for the ferromagnetic
correlations suggested by the Goodenough-Kanamori rules for alternating
orbital order. Without it, the singlet and triplet `virtual' states occurring
in the spin-orbital (`Kugel-Khomskii') exchange processes along
an alternating-orbital bond, illustrated in Fig.~\ref{fig:O_SO_mechanism}, would have the same
energy -- cf. Ref.~\onlinecite{Oles2005} for detailed examples. This would
yield the same amplitude for the two exchange processes and would thus remove all energy gain from a ferromagnetic
alignment relative to the antiferromagnetic (AFM) one.

Despite the fundamental importance of Hund's exchange for the ground-state
ordering, little is known about its signature in dispersive collective orbital excitations
(orbitons). Such excitations were observed in quasi-1D copper
oxides with almost decoupled $S=1/2$ antiferromagnetic
chains~\cite{SchlappaOrbitons2012, Wohlfeld2013, Bisogni2015} as well as in
quasi-2D iridate Sr$_2$IrO$_4$~\cite{Kim:2012cr,Kim:2014ku}. In a Mott-Hubbard insulator, the orbiton moves via
superexchange processes that are rather similar to the ones underlying the
Goodenough-Kanamori rules, see the sketch
Fig.~\ref{fig:O_SO_mechanism}. Interpreting this as an 'orbiton hopping', the situation of an
orbiton moving through an antiferromagnet was then described with
a  minimal  $t$--$J$ model, in perfect analogy to a
hole in the same background~\cite{Wohlfeld2011}. However, this -- quite
successful -- treatment requires the orbiton hoppings on ferro- and
antiferromagnetically aligned bonds to be equal, while Hund's exchange implies that they are not.

The question of Hund's exchange becomes particularly salient in the 1D case, where the
picture of spin-orbital separation was based on the above analogy to a hole:
whereas an orbiton always strongly couples to the elementary spin excitations of an antiferromagnet,
it can effectively separate from the spin excitation (`spinon') in a
similar manner as a `holon' separates from the spinon when a single
hole is introduced into a 1D antiferromagnet~\cite{kim_observation_1996}. The
fate of spin-orbit separation in the presence of Hund's exchange was only
recently addressed theoretically~\cite{heverhagen_spinon_orbiton_2018}.
Fractionalization into spinon and orbiton was predicted to
persist, even though Hund's exchange was concluded to mediate interactions
between them. However, experimental information on the impact of Hund's
exchange on orbiton propagation and spin-orbital separation is so far missing.

In this paper we present a systematic and detailed high-resolution resonant inelastic x-ray scattering (RIXS)
study at the Cu $L_3$ edge on the quasi-1D spin $S=1/2$ antiferromagnetic Heisenberg chain Ca$_2$CuO$_3$.
We assess the importance of Hund's exchange in modeling the experimentally
observed orbital excitations and conclude it to be necessary for a
quantitative description. To this end,
we firstly discuss the experimental methods and present the main RIXS spectra in Sec.~\ref{ExperimentalMethods}.
We start the discussion with analyzing the spin excitations in great detail, see Sec.~\ref{TwoSpinon}. Next, we introduce an appropriate spin-orbital model in Sec.~\ref{TheoreticalModel} and compare the experimental
and theoretical results in Sec.~\ref{Comparison}, paying particular attention to the role of the finite Hund's exchange in obtaining the results
which well-describe the experiment. The paper is summarized in
Sec.~\ref{Conclusions} and supplemented by three
Appendices~\ref{appendix:XRD}-\ref{appendix:theo_model}-\ref{appendix:eff_ham},  in which details about Ca$_2$CuO$_3$ and of the theoretical model are discussed.

\begin{figure*}[htbp]
	\includegraphics[width=1.8\columnwidth]{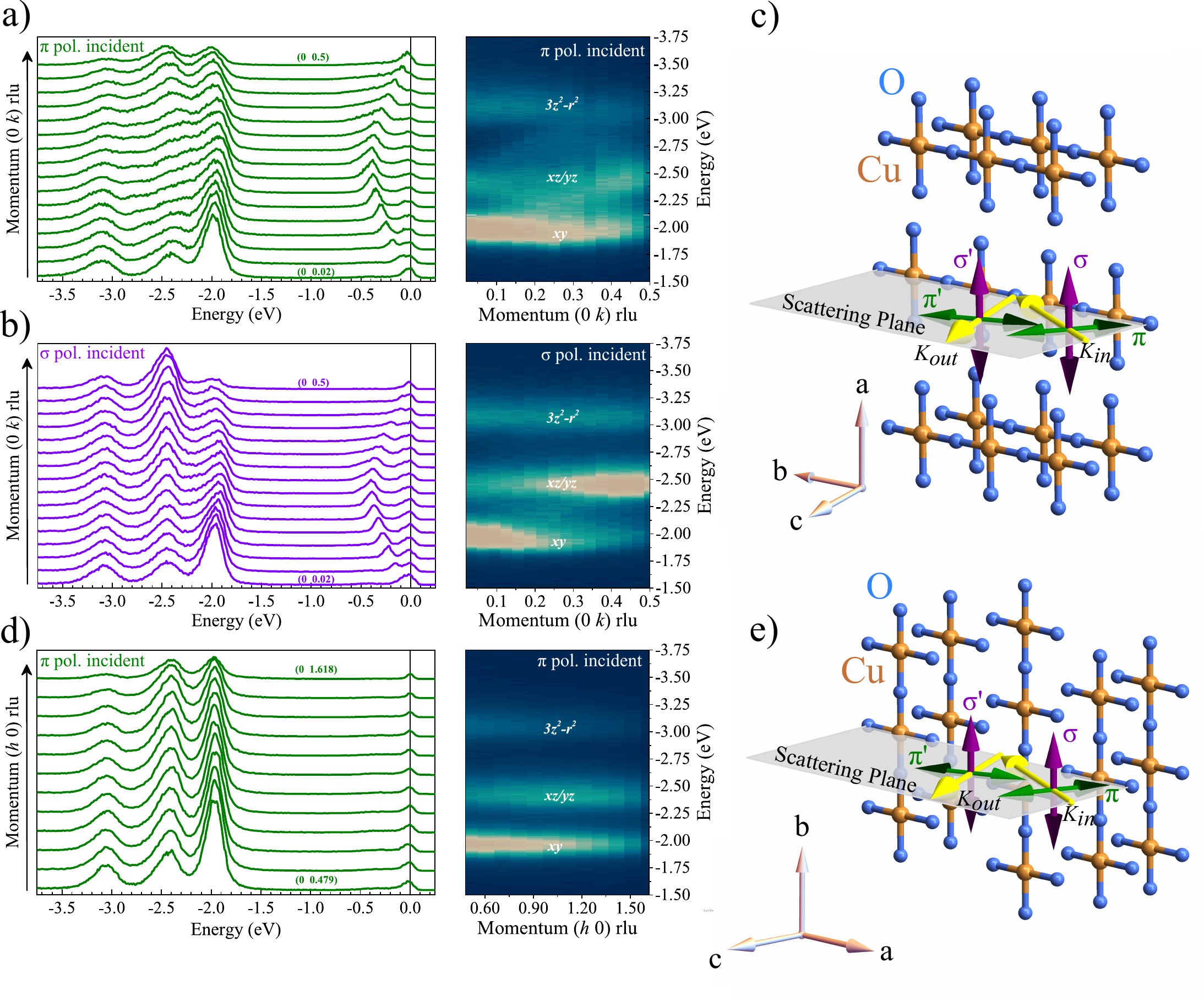}
	\caption{\label{fig:Figure1} Raw RIXS spectra of Ca$_{2}$CuO$_{3}$ measured along the [0 1] direction with $\pi$ (a) and $\sigma$ (b) polarization of the incident light; right panels show a close-up view of the RIXS intensity maps in the $dd$ excitations energy range. c) Sketch of the scattering and experimental geometry adopted to measure the spectra shown in a) and b) together with the crystal structure of CuO$_3$ chains of Ca21. d) RIXS spectra and close-up view of the orbital excitations measured with $\pi$ polarization along the direction perpendicular to the CuO$_3$ chains ([1 0]). e) Experimental geometry adopted for the data presented in panel d). In c) and e) the Ca atoms have been removed for clarity.}
\end{figure*}

\section{Experimental methods}
\label{ExperimentalMethods}

\subsection{Samples}

Together with Sr$_{2}$CuO$_{3}$, Ca$_{2}$CuO$_{3}$ (from now on Ca21) represents one of the best prototype of the quasi-1D spin $\nicefrac{1}{2}$ AFM strongly anisotropic Heisenberg chain\cite{Rosner1997}. The dicalcium cuprate shares the common crystal structure of the better studied Sr$_{2}$CuO$_{3}$\cite{Teske1969,Hjorth1990,Rosner1997} and it is characterized by having quasi-1D CuO$_{3}$ chains of corner-sharing CuO$_{4}$ plaquettes along the crystallographic $b$ axis. Regarding the electronic properties, the strong on-site Coulomb $\emph{U}$ repulsion results in a ground state with one localized hole per Cu ion, located in the 3$d_{x^2-y^2}$ orbital. Because of that, its 3$d$ bands structure shows a Mott-Hubbard gap.
The Ca21 films used for this work were grown by pulsed laser deposition, using a $\lambda$ = 248 nm KrF excimer laser. We chose as the substrate for the film deposition a 5x5 mm$^2$ LaSrAlO$_4$ (LSAO) (1 0 0), since it has a crystal structure (K2NiF$_4$-type) similar to  Ca$_{2}$CuO$_{3}$ and compatible in plane lattice parameters. On LSAO (1 0 0), the Ca21 grows along the $c$-axis and the CuO$_{3}$ 1D chain lies in the $ab$ plane along the $b$-axis (see Fig.\,\ref{fig:Figure1}(c)), with chain oxygen at corner sharing CuO$_4$ plaquettes. The Ca21 target for the film deposition was prepared by standard solid state reaction: stoichiometric mixtures of high-purity CaCO$_3$ and CuO powders were calcined at about 800$^\circ$C in air for 20 hours, then pressed to form a disk and heated in air at 950 $^\circ$C for 24 h. The distance between LSAO substrate and the target was 2.5 cm. The substrate holder was kept at  T = 600 $^\circ$C during the deposition at an oxygen pressure of 0.1 mbar, and cooled down to room temperature at the same pressure. High quality, totally detwinned, $c$-axis oriented films were grown with thickness about 30 nm (see Appendix \ref{appendix:XRD}).

\subsection{RIXS Measurements}
\label{RIXS}

\begin{figure}[htbp]
	\includegraphics[width=1 \columnwidth]{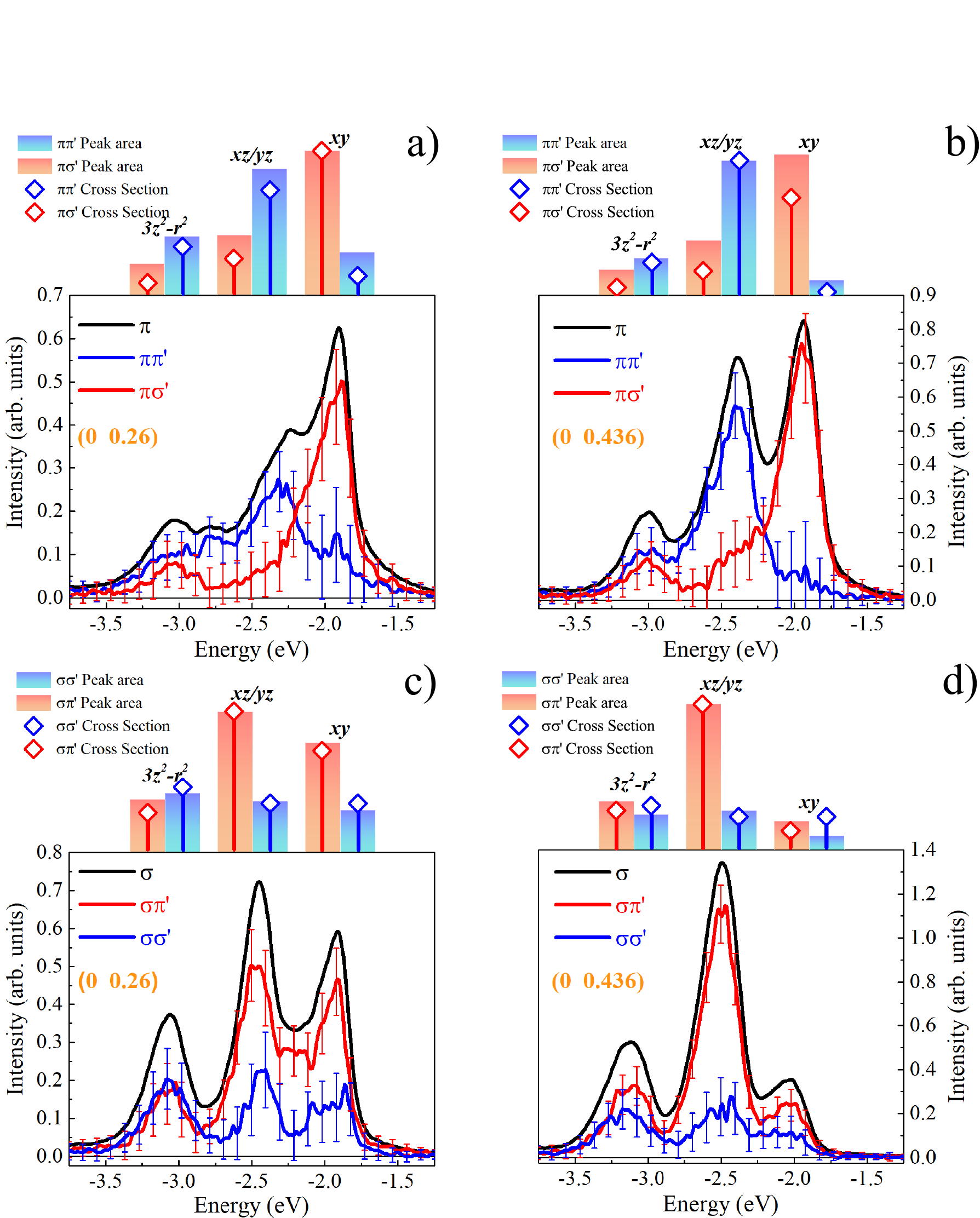}
	\caption{\label{fig:Figure2} Polarimetric RIXS spectra measured with incident $\sigma$ (top panels) and $\pi$ (bottom panels) polarization at two selected transferred momenta (0.26 and 0.436 r.l.u.). On top of each spectrum we report the comparison between absolute values of the polarization-resolved RIXS cross section calculations and the peak areas for each $dd$ excitation. To help the reader, we use the same color code (blue for the non crossed $\pi\pi^\prime$ and $\sigma\sigma^\prime$ and red for the crossed $\pi\sigma^\prime$ and $\sigma\pi^\prime$ channels) for both the experimental data and the histograms. }
\end{figure}

RIXS\cite{AmentRev2011} measurements were carried out at the beamline ID32\cite{ID32Beamline} of the ESRF - the European Synchrotron. The energy of the incident beam was tuned to the maximum of the Cu $L_{3}$ absorption peak ($\sim \SI{931}{\electronvolt}$) in order to fulfill the resonance condition. The polarization of the incident light was set either parallel ($\pi$) or perpendicular ($\sigma$) to the scattering plane. For all the measurements discussed below, the scattering angle 2$\theta$ was fixed at \SI{149.5}{\degree} in order to maximize the momentum transfer. The sample temperature was kept fixed at \SI{20}{\kelvin}. To explore how the dimensionality of Ca21 affects the RIXS spectra, we were able to rotate by \SI{90}{\degree} the azimuthal angle so that either the $bc$ or $ac$ planes of the sample were parallel to the scattering plane, as shown in Fig.\,\ref{fig:Figure1} c) and e), respectively. Once the scattering plane is defined, by rotating the angle perpendicular to it ($\theta$) we were able to change the in-plane transferred momentum $q_\parallel$, defined as the projection of the momentum transfer onto the CuO$_{2}$ layers. From now on we will refer to the transferred momentum values in terms of reciprocal lattice units (r.l.u.) 2$\pi/a$, 2$\pi/b$ and 2$\pi/c$. We acquired RIXS spectra along the two high-symmetry directions [0 1] (Fig.\,\ref{fig:Figure1} a-c) and [1 0] (Fig.\,\ref{fig:Figure1} d-e) in the first Brillouin zone. Note that, due to the orthorombicity of the
crystal structure, these two directions are not equivalent. In fact, when the scattering occurs in the $bc$ plane, the CuO$_3$ chains lie in the scattering plane; on the other hand, if the scattering plane is parallel to the $ac$ plane, the chains are perpendicular to it. These considerations on the different geometries have an important impact and significant repercussions on the physics of 1D systems, as we will explain in more details in Sec.\,\ref{ResultsandDiscussions}.

The spectra displayed in the waterfall plots of Fig.\,\ref{fig:Figure1} can be decomposed into several features. The peaks at $\sim$ 0 energy loss represent the quasi-elastic scattering (which includes diffuse elastic and phonons); up to \SI{-0.5}{\electronvolt} all the spectra reveal the presence of dispersing spin excitations and between -1.5 and \SI{-3.5}{\electronvolt} the spectra are dominated by orbital excitations, which correspond to the final states with the Cu hole in the other 3$d$ orbitals. The assignment of each orbital energy and symmetry has been done by fitting the experimental spectra with only 3 peaks whose relative intensity was assigned following the  RIXS cross sections calculated in a pure ionic picture \cite{ddMoretti}.

We also exploited the unique capability of the {ERIXS} (European-RIXS) spectrometer, to perform polarization-resolved RIXS measurements\cite{ID32Beamline,Fumagalli2019Polarimeter}, disentangling the two linearly polarized channels ($\pi^\prime$ and $\sigma^\prime$) of the scattered light. The possibility of disentangling the polarization of the scattered light gives us valuable insights on the nature of the various spectral features: recently, it has been demonstrated that this method can be useful to distinguish the various orbital\cite{Fumagalli2019Polarimeter} and the low-energy excitations\cite{LucioPolarimeter,PengNatMat2018,HeptingNature2018,SilvaNetoMM_Polarimeter} in different cuprate families. Eventually, we combined the data taken with and without the polarimeter to assign all the final states as labelled in the colormaps of Fig. \ref{fig:Figure1}. In  Fig.\,\ref{fig:Figure2} we show the polarimetric raw RIXS data of Ca21 measured with both incident $\pi$ (bottom panels) and $\sigma$ (top panels) polarizations at two distinct transferred momenta values (0.26 and 0.436 r.l.u.). The decomposed outgoing polarization-dependent channels and the experimental error bars have been obtained following the procedure reported in Refs.\,\onlinecite{LucioPolarimeter,Fumagalli2019Polarimeter}. On top of each spectrum we report the comparison between the absolute values of the polarization-resolved RIXS cross sections calculated within the single ion model as a function of the incident/scattered polarizations and the peak areas for each orbital excitation. The results are shown in the form of histograms keeping the same color code of the experimental data. We remark the good agreement between the calculations and the experimental data. Most importantly, we find that the region showing the larger dispersion in Fig.\,\ref{fig:Figure2}(a) obtained with $\pi$ polarization has a predominant $d_{xz/yz}$ character. The results obtained by combining the polarimetric data and the calculations will be used in the detailed analysis presented throughout this paper.

The overall energy resolution was $\sim\SI{60}{\milli\electronvolt}$ for the unpolarized spectra, while we relaxed it to $\sim\SI{80}{\milli\electronvolt}$ for the polarimetric measurements in order to maintain an acceptable count rate. Each RIXS spectrum shown in Fig.\,\ref{fig:Figure1} has been acquired in 15 min. In the case of polarization-resolved measurement, the acquisition time was increased to 60 min in order to get good statistics due to the lower efficiency of the polarimeter\cite{Fumagalli2019Polarimeter}.

\section{Results and Discussions}
\label{ResultsandDiscussions}

\subsection{Two-spinon continuum}
\label{TwoSpinon}

Spin excitations in quasi-1D systems have been intensively studied in the last decades due to their importance in the realization of the 1D Heisenberg AFM model. The anisotropic AFM interaction of Ca21 comes from the Cu atoms within the chains, along the $b$
axis. The coupling along the other crystallographic directions is negligible, making Ca21 a 1D magnetic system with a N\'{e}el temperature of only \SI{9}{\kelvin}~\cite{Rosner1997} despite the large nearest neighbor superexchange interaction along the chains. In fact, RIXS
spectra measured along the [1 0] direction do not show any dispersive spin excitations.

\begin{figure}[htbp]
	\includegraphics[width=1\columnwidth]{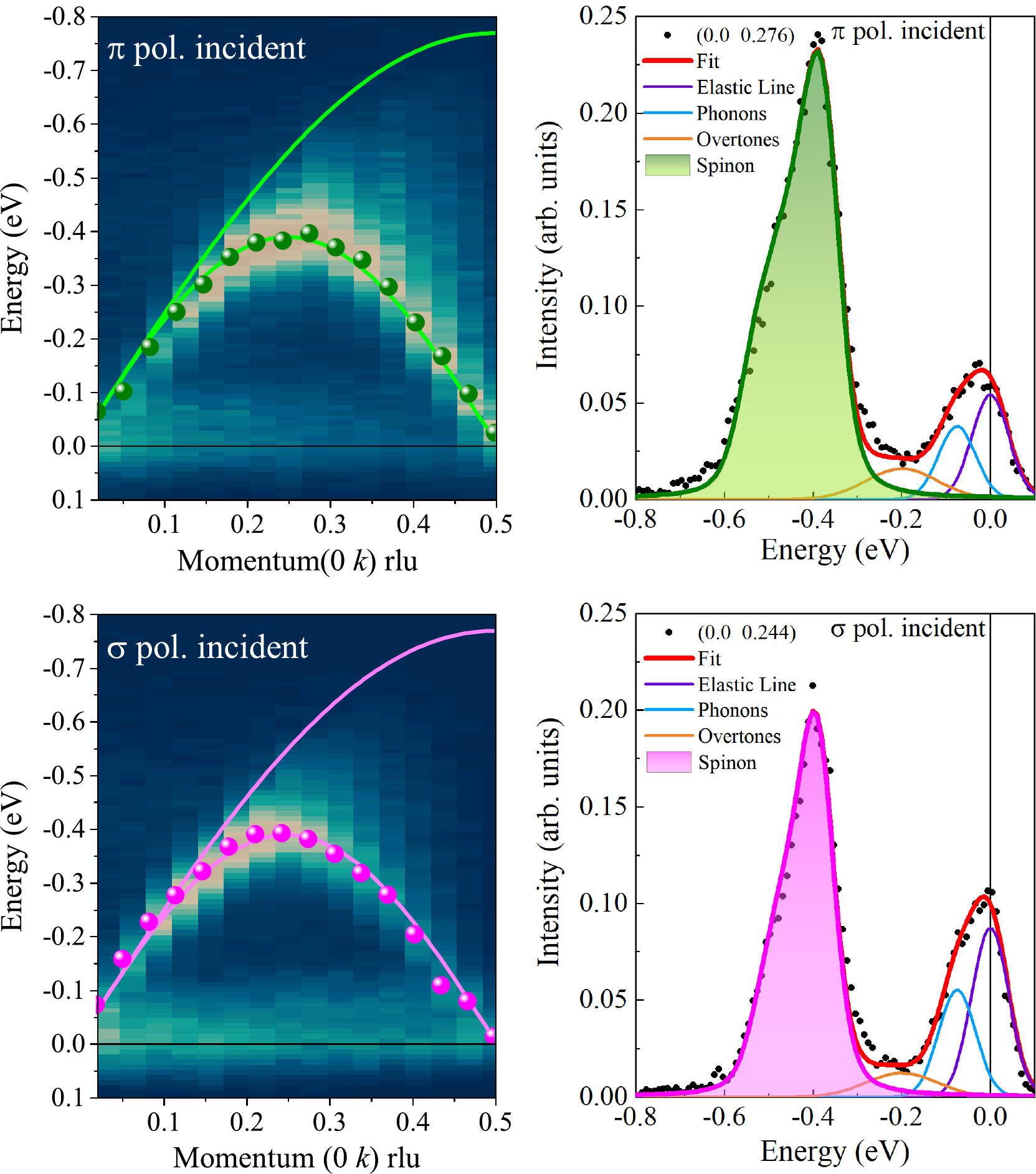}
	\caption{\label{fig:Figure3} Momentum transfer dependence of the magnetic two-spinon continuum of Ca$_{2}$CuO$_{3}$ taken with $\pi$ (top left panel) and $\sigma$ (bottom left panel) polarization measured along the [0 1] high-symmetry direction. The results of the fittings are represented by dots (energy position of the spin excitations) while the solid lines are related to the approximate two-spinon continuum boundary.
	The right panels show the low energy RIXS spectra at two specific transferred momenta, which were fitted by assuming the onset of four distinct features:
	the quasi-elastic peak (purple), a generic phonons contribution (blue), its overtones (orange) and the spinon (green/magenta for $\pi$/$\sigma$ polarization). Whereas the first three features were fitted with Lorentzian line-shape, the latter was fitted with the approximated two-spinon dynamical structure factor $S(\textbf{q},\omega)$ and convoluted with a gaussian lineshape~\cite{karbach1996} in order to take into account the overall experimental energy resolution.}	
\end{figure}

A clear signature of the 1D nature of Ca21 comes directly from the momentum dependence of spin excitations along the [0 1] direction. Indeed, in these systems the elementary magnetic excitations with $S = 1$ fractionalize into the two-spinon continuum carrying a
spin quantum number of $S=\nicefrac{1}{2}$~\cite{karbach1996}. In the left panels of Fig.\,\ref{fig:Figure3} we show the Ca21 RIXS intensity maps (between 0 and \SI{0.8}{\electronvolt} of energy loss) measured with both $\pi$ (top left) and $\sigma$ (bottom left) polarization which display strong dispersing spin excitations characterized by two different periodicities in reciprocal space~\cite{Gianmarchi2004,Caux2006,Caux2008,Lake2005,Lake2009,Mourigal2013}: $\pi$ for the lower edge of the two-spinon continuum, 2$\pi$ for the upper edge. In the intensity maps we show the extracted energy positions of the main two-spinon peak from a preliminary fitting procedure (dots). In order to determine the magnetic superexchange constant ($J_\textrm{SE}$), we assume that RIXS probes the spin dynamical structure factor $S(\textbf{q},\omega)$. This approximation has been intensively used through the years in RIXS~\cite{SchlappaOrbitons2012,Glawion2011,AmentRev2011,Kourtis2012,Forte2011,Schlappa2018} experiments on a large number of quasi-1D antiferromagnets. Here we used the approximated expression of the two-spinon dynamical structure factor reported in Ref.~\onlinecite{karbach1996} to obtain a value of the superexchange coupling $J_\textrm{SE} = \SI{0.24}{\electronvolt}$ for the present case of Ca21, which is not only close to the value found for Sr$_{2}$CuO$_{3}$~\cite{Walters2009,SchlappaOrbitons2012} but also to the one reported for Ca21 \cite{Suzuura1996}.
The results are represented by the continuous lines shown in the left panels of Fig.\,\ref{fig:Figure3}.
Furthermore, thanks to the good quality of our RIXS spectra, we can directly compare the RIXS line shapes with the approximated expression of the two-spinon dynamical structure factor $S(\textbf{q},\omega)$ of the $S=1/2$ Heisenberg chain~\cite{karbach1996}.

Regarding the line shape, the obtained intensity of our fitting is given by the following formula
\begin{equation} \label{equ:lineshape_spinon}
I(\textbf{q},\omega) = S(\textbf{q},\omega)*G(\omega),
\end{equation}
\noindent where * is the convolution and
\begin{equation}
S(k,\omega) = \frac{\Theta(\omega - \omega_L(\textbf{q})\Theta(\omega_U(\textbf{q}) - \omega)}{\sqrt{\omega^2 - \omega^2_L(\textbf{q})}}
\end{equation}
\noindent is the approximate expression for the two-spinon dynamic structure factor reported in Ref. \onlinecite{karbach1996}. Here $\Theta$ represents the Heaviside step function, while $\omega_L(\textbf{q}) = \nicefrac{\pi}{2}\sin({\textbf{q}})$ and $\omega_U(\textbf{q}) = \pi\sin({{\nicefrac{\textbf{q}}{2}}})$. Finally, the overall experimental energy resolution is taken into account by convolving $S(\textbf{q},\omega)$ with the following gaussian line shape
\begin{equation}
G(\omega) = \exp(-(\omega-\omega_0)^2/2\sigma^2) / (\sigma \sqrt{2 \pi})
\end{equation}
\noindent with $\sigma$= 55 meV (overall energy resolution) and $\omega_0$ is the peak energy.
In the two right panels of Fig.\,\ref{fig:Figure3} we show the fitting results at two transferred momenta values, which have been chosen by considering the fact that here the two-spinon continuum is well separated from the other spectral features. We would like to underline that the two-spinon contribution to the spin dynamical structure factor is of the order of 73\% of the total (theoretical) spectral weight. Thus, the amplitude of the theoretically
calculated spin dynamical structure factor has been multiplied by an
overall scaling factor (the same for all momenta), in order to match the
theoretical intensity with the experimental data.

\subsection{Theoretical model for $dd$-excitations}
\label{TheoreticalModel}

$dd$ excitations in 1D systems feature the fractionalization of spin and orbital degrees of freedom.
To model the fractionalization\cite, previous studies used an effective $t$-$J$ model, which accurately described the experimental spectra.
We go one step further and include the effect of Hund's coupling, which should be finite in realistic materials.
Note that finite Hund's coupling preserves fractionalization\cite{heverhagen_spinon_orbiton_2018,Daghofer2008}, but hinders the description with an effective $t$-$J$ model and might lead to a non-negligible interaction between orbitons and spinons.

To verify the spinon-orbiton separation, we employ a Kugel-Khomskii-type Hamiltonian~\cite{Kugel1982}, which can be written in the following general form:
\begin{align} \label{equ:kk_ham}
H = 2 \sum^{}_{\langle i ,j\rangle} \left( \vec S_i
\cdot \vec S_j + \frac{1}{4}  \right) A(T^\beta_i,T^\alpha_j) \nonumber \\
+ \sum^{}_{\langle i,j\rangle}K(T^{\beta}_{i},T^{\alpha}_j)
&+ \Delta_\textrm{CF}\sum_{i} T^z_{i}\; .
\end{align}
Here $\vec{S}_i$ describes a spin $S=1/2$ at site $i$ and $T^\alpha_i$ is a pseudo-spin $T=1/2$
with component $\alpha\in \{x,y,z\}$ to describe the orbital degree of freedom between the ground state $d_{x^2-y^2}$ and one of the excited orbitals $d_{xy}$, $d_{xz}$ and $d_{yz}$ (note that the `nondispersive' $d_{3z^2-r^2}$ orbital is not taken into account in the analysis).
Bonds $\langle i ,j\rangle$  take nearest neighbors into account and $\Delta_\textrm{CF}$ is the crystal field splitting.
Operators $A(T^\beta_i,T^\alpha_j)$ and $K(T^{\beta}_{i},T^{\alpha}_j)$ only depend on the orbital degrees of freedom and account for onsite repulsion $U$, Hund's coupling $J_\textrm{H}$ and nearest-neighbor hopping $t_1$ and $t_2$ for the ground state and the excited orbital, respectively (see Appendix~\ref{appendix:theo_model} for details).
Note that for the ferro-orbital ground state, Hamiltonian \eqref{equ:kk_ham} reduces to the Heisenberg model, which describes the spin excitations discussed in the previous section.

To model the orbital excitation in the RIXS spectrum, we calculate the orbital spectral functions $O(q,\omega)$ and the spin-orbital $SO(q,\omega)$ with exact diagonalization and cluster perturbation theory for each orbital separately.
The exact diagonalization results are broadened by a Lorentzian line shape of 120 meV FWHM to account for finite size effects (for details see Appendix~\ref{appendix:theo_model}).
For direct comparison with the experiment, they have to be multiplied by the RIXS `matrix elements' that can easily be obtained using the so-called fast collision approximation to the Kramers-Heisenberg formula for RIXS (for more details, see Refs. \onlinecite{AmentRev2011,ddMoretti,Fumagalli2019Polarimeter}).
Finally, the results are convoluted with a Gaussian function ($\text{FWHM}=60\, \text{meV}$) to account for experimental resolution.

Even in the presence of Hund's coupling, the spinon-orbiton fractionalization
is preserved \cite{heverhagen_spinon_orbiton_2018}, although it mediates an interaction between orbiton and spinon that can obscure signatures of spin-orbit separation if it becomes too strong.
In fact, for finite Hund's exchange the hopping of an orbiton is modified, as shown in Fig.\,\ref{fig:O_SO_mechanism}, by the distinct
superexchange processes for anti-parallel and aligned spins.
For anti-parallel spins an excited electron can move with superexchange constant $t_1t_2/(U-2J_\textrm{H})$, which increases to $t_1t_2/(U-3J_\textrm{H})$ in the case of parallel spins.
Hence, orbital (\emph{O}) and spin-orbital (\emph{SO}) excitations are  distinguishable for non-zero $J_\textrm{H}$ and need to be considered separately.
Additionally, a larger superexchange constant also increases the
bandwidth of the excitations ($W$) as well as its dispersion. The
latter is directly connected to the
orbiton velocity \cite{Wohlfeld2011}, which means that a larger $J_H$ increases
the velocity of the orbiton. In fact, the orbiton velocity is defined as
\begin{equation}
v_{\rm orb} = \left.\frac{ d \varepsilon_{\rm orb}}{ dk} \right\rvert_{k=0} = 2 \tilde{t},
\end{equation}
for the (bare) orbiton dispersion, which is given by $\varepsilon_{\rm orb} (k) = -2 \tilde{t} \sin (k)$ [see Ref. \onlinecite{Wohlfeld2011} or Appendix \ref{appendix:eff_ham}] with the orbiton hopping $\tilde{t} = 2 t_1 t_2/(U'(1-J_H/U'))$ (see Appendix \ref{appendix:eff_ham} for details).
Note that this intuitive picture is supported by numerical calculations \cite{heverhagen_spinon_orbiton_2018}.

Using this picture, we derived the approximate analytic relation between $J_H$ and $W$ (see Appendix \ref{appendix:eff_ham})
\begin{align} \label{equ:jhund}
J_\textrm{H} = \left( U- \frac{8t_1t_2}{W} \right) \frac{1}{3} ,
\end{align}
which makes it possible to calculate $J_\textrm{H}$ from the bandwidth of the excitation.
This implies that a theory for spinon-orbiton separation which ignores Hund's coupling always overestimates the hopping constant of the excited orbital $t_2$.

\subsection{Comparison between experiment and theory}
\label{Comparison}

In Fig.\,\ref{fig:Figure1} we show the measured RIXS spectra of Ca21 in different experimental configurations. As mentioned above, the quasi-1D nature of Ca21 has strong repercussions on the dispersion of the spectral features seen by RIXS. This is already observed in the absence of spin excitations when spectra are measured with the CuO$_3$ chains oriented perpendicularly to the scattering plane. In this configuration, we do not see any dispersion in the $dd$ excitations energy range (panel d of Fig.\,\ref{fig:Figure1}). Therefore, let us from now on focus solely on the geometry in which the chains are parallel to the scattering plane [panels (a-c) of Fig.\,\ref{fig:Figure1}]. This configuration is similar to the one reported by Bisogni \emph{et al.}~\cite{Bisogni2015}, where in the quasi-1D AFM spin-ladder CaCu$_2$O$_3$, the spin-orbital separation occurs along the $a$ direction in the $d_{xz}$ orbital channel.

\begin{figure}[htbp]
	\includegraphics[width=1\columnwidth]{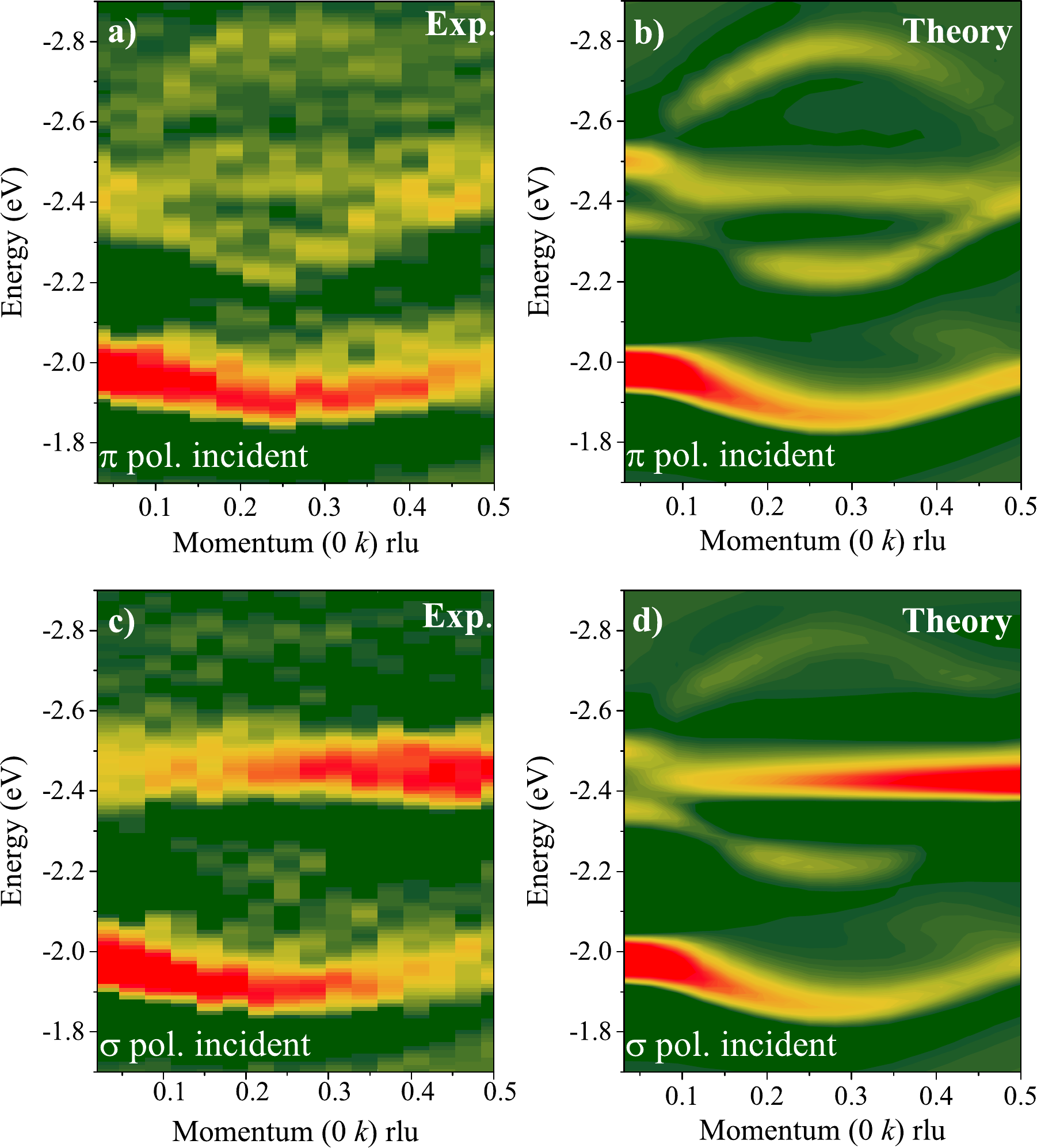}
	\caption{\label{fig:Figure5} Experimental (a,c) and theoretical (b,d) `best fit' second derivative RIXS intensity maps in the orbital excitations energy range. Top (bottom) panels show results for the $\pi$ ($\sigma$) incoming polarization, respectively.}
\end{figure}

To explain the observed orbital spectra, with their peculiar dispersion relations, we use the theoretical model described in detail in Sec.~\ref{TheoreticalModel}. We choose the parameters of the model
in two steps. First, we start with the parameter related to the $d_{x^2-y^2}$ orbital -- the hopping constant $t_1$ between the two nearest neighbor $d_{x^2-y^2}$ orbitals.
The latter could be calculated from to the superexchange constant $J_\textrm{SE} = 4t^2_1/U = 0.24 $ eV,  as obtained from the RIXS spin spectrum in Sec.~\ref{TwoSpinon}.
Assuming that typically for the cuprates the on-site Coulomb repulsion $U= 8 t_1$~\cite{jia_persistent_2014} we get that $t_1= \SI{0.49}{\electronvolt}$ and $U= \SI{3.92}{\electronvolt}$.

We obtain the remaining model parameters ($t_2$ describing the hopping between the excited orbitals, the Hund's coupling $J_\textrm{H}$ and the on-site crystal field energies $\Delta_\textrm{CF}$ of the excited orbitals, i.e., $d_{xy}$, $d_{yz}$, $d_{xz}$) by directly comparing the {\it second derivatives} of the theoretical results and of the experimental spectra, see Fig.\,\ref{fig:Figure5}.
The results from the `best fit' of the theoretical model to the experimental data are given in Table\,\ref{table:Parameters}.
As can be seen in Fig.\,\ref{fig:Figure5} the experimental and theoretical second derivative maps are in very good agreement.
Let us note in passing that the second derivative curves easily track the peak position of the different spectral features, making the dispersion of the orbital excitations more evident without any type of fitting -- in fact, such a method has been previously used in the analysis of RIXS data to disentangle dispersive orbital excitations from the particle-hole continuum in Sr$_2$IrO$_4$\cite{Kim2012}.

We note that the dispersion of orbital excitation is evident in the spectra measured with $\pi$ polarization and hardly visible with $\sigma$ (see Fig. \ref{fig:Figure1}a,b, and also Ref.~\onlinecite{SchlappaOrbitons2012}). This observation might be naively interpreted as a polarization-dependence of the spin-orbital fractionalization phenomenon. Actually this apparent polarization-dependentence is not happening in for the pure spin excitations in the low energy scale, as widely discussed in Sec.\,\ref{TwoSpinon}: the fractionalized spinon excitations show the same momentum dependence for both $\pi$ and $\sigma$ incoming light. To unravel this puzzle we take a closer look at the second derivative maps in the case of $\sigma$ incident polarization shown in the bottom panels (c-d) of Fig.\,\ref{fig:Figure5}. They reveal the presence of a dispersion even for $\sigma$ incoming polarization, although it is very weak. What happens is that with $\sigma$ polarization the non-dispersing $d_{xz}$ RIXS cross section is much larger than that of the dispersing $d_{yz}$. On the contrary, with the $\pi$ polarization the $d_{yz}$ excitation provides most of the RIXS signal, leading to an apparent absence of overall dispersion.

\begin{table}
	\begin{tabular}{lccc}
		\hline\hline
		&$xy$ &$xz$&$yz$ \\ \hline
		$J_\textrm{H}$ (in units of $t_1$)& $0.8 (\pm 0.2)$& $1.2 (\pm 0.2)$ & $1.2 (\pm 0.2)$\\
		$t_2$ (in units of $t_1$)& $0.5 (\pm 0.1)$& $0 (\pm 0.2)$ & $0.7 (\pm 0.1)$\\
		$\Delta_\textrm{CF}$ (eV) &1$.87 (\pm 0.03) $& $2.32 (\pm 0.03) $& $2.45 (\pm 0.03)$\\
		%$2\Gamma_L$ (eV)& 0.06 & 0.06 & 0.06\\
		\hline\hline
	\end{tabular}
	\caption{`Best fit' microscopic parameters for the Kugel-Khomskii Hamiltonian \eqref{equ:kk_ham}. For definitions see main text and Appendix.
	Errors are estimated by comparing the second derivatives of the experiment with the theoretical spectra (see Appendix \ref{appendix:theo_model} for details).
	   }
	\label{table:Parameters}
\end{table}

\begin{figure}[t]
	\includegraphics[width=1\columnwidth]{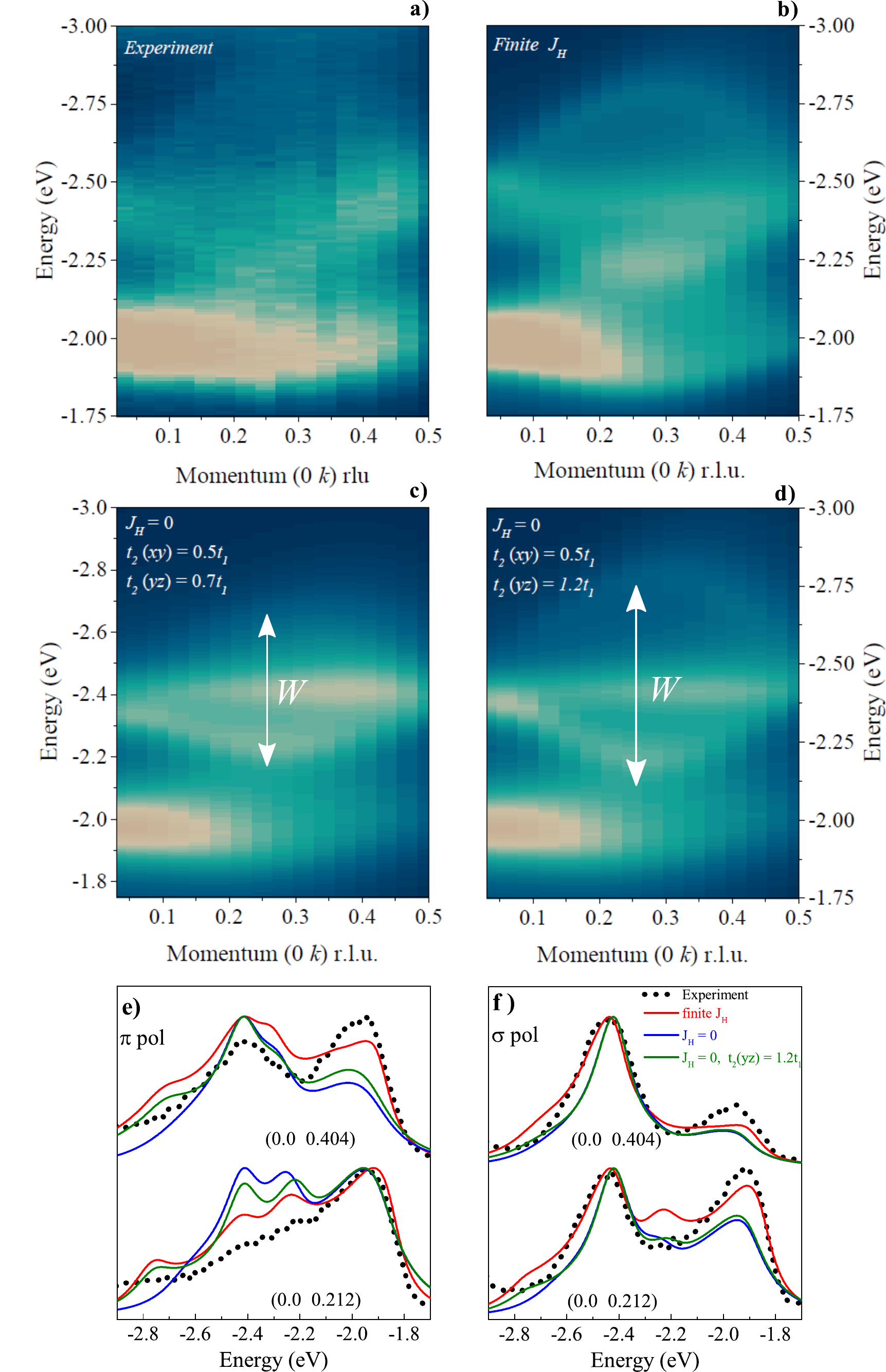}
	\caption{\label{fig:Figure6} Comparison between the experimental and theoretical RIXS spectra [(a) $vs$ (b-d) panels] and the line spectra (e-f): panel (a) shows a `zoom-in' into the orbital part of the experimental RIXS spectra of Ca21; panel (b) shows the best theoretical description (`best fit') of the experimental data with the model parameters listed in Table\,\ref{table:Parameters}; panels (c,d) show results for Hund's coupling $J_H=0$ with two different values of the orbiton hopping parameters $t_1, t_2$ ($W$ indicates the bandwidth of the excitation, which is larger for an increased $t_2$); panels (e-f) show experimental as well as the theoretical line spectra, calculated with the model parameters as used in panels (b-d) and presented at two selected momentum transfers. The spectra in panels (a) and (e) are measured with $\pi$ incident polarization, while the spectra in panel (f) are measured with $\sigma$ incident polarization.}
\end{figure}

The theoretical RIXS spectra calculated with the `best fit' model parameters from Table~\ref{table:Parameters} are shown in Figs.~\ref{fig:Figure6}(b, e, f). Indeed, as already suggested by the second derivative maps of Fig.~\ref{fig:Figure5}, one can observe a very good agreement with the experimental RIXS spectrum shown, e.g., in Fig.~\ref{fig:Figure6}(a):
{\it First}, the dispersion of the lowest energy excitation ($d_{xy}$) is well reproduced in both the cases with $\pi$ and $\sigma$ incident light polarization.
Its momentum dependence is characterized by \nicefrac{1}{2} r.l.u. periodicity in reciprocal space, as also shown in Sr$_{2}$CuO$_{3}$\cite{SchlappaOrbitons2012}.
As in previous studies for Sr$_{2}$CuO$_{3}$, the $d_{xz}$ and $d_{yz}$ energies are split by  $\sim\SI{150}{\milli\electronvolt}$ \cite{Wohlfeld2013}.
{\it Second}, regarding the $d_{xz}$ orbital excitation, assuming $t_2=0$ (due to the negligible overlap between nearest-neighbor $d_{xz}$ orbitals along the $y$ direction of the chain) results in a non-dispersive excitation.
{\it Third}, the spin-orbital fractionalization is really very well-visible for the $d_{yz}$ excitation.
Thus, the momentum dependence of the $d_{yz}$ dispersion shows two strongly dispersive branches giving a characteristic oval shape~\cite{Wohlfeld2011, Wohlfeld2013}.
This, as discussed in detail in Appendix \ref{appendix:eff_ham}, allows us to extract the bandwidth of the excitation $W\sim \SI{0.6}{\electronvolt}$.
The bandwidth can then be used in Eq.~\eqref{equ:bandwidth} to obtain the Hund's coupling to be of the order of $\simeq 1.14 t_1$, which is in good agreement with the above-discussed `best fit' procedure (compare Tab. \ref{table:Parameters}). Overall the values of Hund's coupling are in good agreement with the ones typically assumed for another quasi-1D cuprate -- Sr$_2$CuO$_3$\cite{Wohlfeld2013}.

Finally, we would like to stress that such a good agreement between the theoretical model (with the parameters from Table~\ref{table:Parameters}) and the experimental result cannot be achieved without a finite value of the Hund's coupling,~cf.~Fig.~\ref{fig:Figure6}(c-d). For instance, if $J_H=0$ and we use the same values of $t_2$  as reported in Table\,\ref{table:Parameters} for all the three orbitals considered we do not get a good agreement with the experimental data. It is then only when a much larger value of the hopping parameter $t_2$ ($t_2>t_1$) is assumed [Fig.~\ref{fig:Figure6} (d)] that we obtain a satisfactory agreement with the experiment. This demonstrates that fitting the experimental results with the theoretical model with $J_H=0$ requires a non-physical, i.e., far too large, hopping $t_2$ -- this is due to the fact that the hopping $t_2$ of the excited orbital should not be larger than the hopping related to the ground state orbitals (since otherwise the hybridisation would lower the energy of the excited orbital below that of the ground state orbital).
Hence, the Hund's coupling is necessary to increase the dispersion of the excitation, which, as noted in Sec.~\ref{TheoreticalModel}, increases the velocity of the orbiton.
Moreover, a detailed comparison of spectral line shapes at selected transferred momenta, see Fig.\,\ref{fig:Figure6}(e-f), shows that the fit with a finite (and realistic) values of Hund's rule coupling provides a closer agreement with experiment than the fit relying on (less realistic) large $t_2$. The remaining differences between theory and experiment may be due to some charge-transfer excitations contributing to the background or to processes not included in our model, e.g., coupling to the lattice and Jahn-Teller effect or longer-range spin and spin-orbital exchange processes.

\section{Conclusions}
\label{Conclusions}
We have studied in detail the RIXS response at the Cu $L$ edge of the quasi-1D antiferromagnet Ca$_2$CuO$_3$.
We observed a clear signature of the dispersive spin and orbital  3$d_{xy}$ and 3$d_{yz}$ excitations, at first sight qualitatively resembling
the ones recently found in other quasi-1D copper oxides~\cite{SchlappaOrbitons2012, Wohlfeld2013, Bisogni2015}.
Thus, we can first conclude that also in this case it is correct to interpret the spin (orbital) excitations in terms of the onset of the two-spinon continuum (spin-orbital separation).

Next, we performed a detailed quantitative analysis of the experimental data. First, in addition to the previous studies, we exploit two distinct incident photon polarizations in our analysis. Moreover, for both polarizations we successfully reproduced the experimental RIXS spectra using a theoretical model based on the well-established Kugel-Khomskii Hamiltonian. Specifically, the crucial finding of this paper is that a detailed modelling of the orbital spectrum
requires a finite Hund's exchange $J_{\rm H}$ in the Hamiltonian. Specifically, for  Ca$_2$CuO$_3$ we obtain that
a moderate, and {\it realistic}, value of $J_{\rm H} \simeq 0.5 $eV best explains the experimental data. Here the main role of the Hund's exchange is
to increase the velocity of the orbiton so that the theoretically predicted one matches well with the experimentally observed one.
We note that the assumed value of the Hund's exchange is relatively small with respect to the calculated spin-orbital exchange constants, so that the interaction
between orbitons and spinons is not that large and thus the spin-orbital separation picture can still approximately describe the physics present here, cf. Ref.~\onlinecite{heverhagen_spinon_orbiton_2018} for details.
Altogether, this means that the Hund's exchange plays a vital role in the propagation of an orbiton.

\section*{Acknowledgements}
We thank Daniel Chrastina for the support with the XRD characterization of the samples. RIXS measurements were performed at the beam line ID32 of the European Synchrotron Radiation Facility (ESRF) with the ERIXS spectrometer. R.F, M.M.S. and G.G. acknowledge support by Fondazione CARIPLO and Regione Lombardia under project ``ERC-P-ReXS'' No. 2016-0790 and by MIUR under project PRIN2017 ``Quantum-2D'' ID 2017Z8TS5B. K.W. kindly acknowledges the support by the Narodowe Centrum Nauki (NCN, Poland) under Projects Nos. 2016/22/E/ST3/00560 and 2016/23/B/ST3/00839.
M.D. acknowledges support via the Center for Integrated Quantum Science and Technology (IQST). R.A. was supported by the Swedish Research Council (VR) under the project ``Evolution of nanoscale charge order in superconducting
YBCO nanostructures (2017-00382)''.
\label{Acknowledgements}

\section*{Appendix}
\appendix
\section{Structural properties of the Ca$_2$CuO$_3$ thin films}
\label{appendix:XRD}

The structural properties of the Ca$_2$CuO$_3$ (Ca21) thin films have been determined by X-Ray Diffraction (XRD) analysis, using a Panalytical X'Pert PRO Materials Research 4-axis diffractometer. The beam from the Cu X-ray tube passes through a 2-bounce Ge220 monochromator, which includes a mirror; the diffracted beam is  detected either by a rocking curve attachment or by a 3-bounce symmetric analyzer crystal, used respectively for low-resolution and high-resolution $\omega$-2$\theta$ scans and reciprocal space maps. The XRD characterization presented in the following is focused on the same sample [30 nm thick Ca21 film on (1~0~0) oriented LaSrAlO$_4$ (LSAO) substrate] on which all the RIXS measurements presented in the paper have been taken. We have however tested that the results are reproducible among several samples of the same kind.

\begin{figure}[htbp]
\includegraphics[width=1\columnwidth]{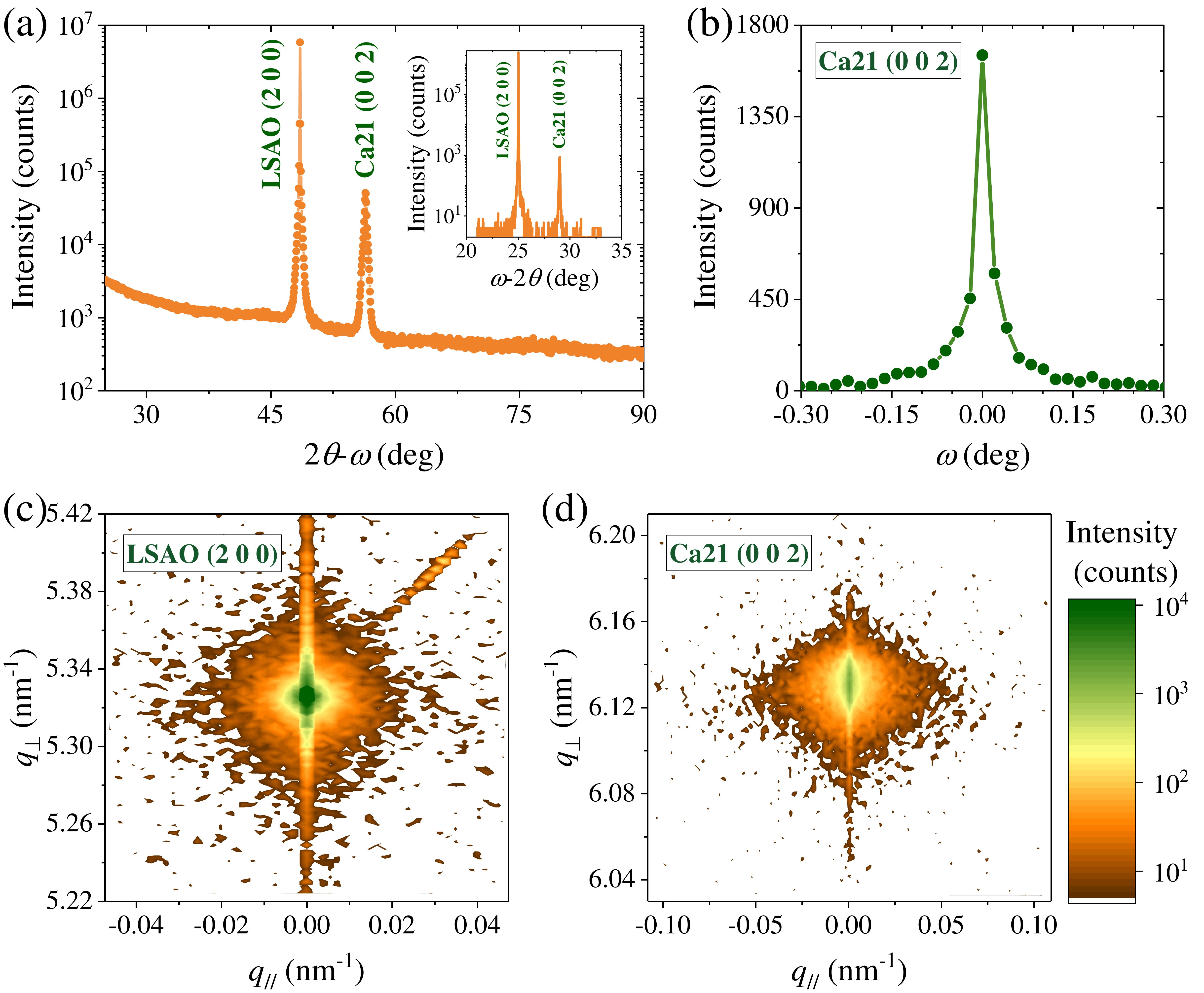}
\caption{Symmetric XRD measurements on the Ca$_2$CuO$_3$ (30 nm)/LaSrAlO$_4$ (1~0~0). (a) The low resolution 2$\theta$-$\omega$ scan shows that, beside the substrate reflection, only the Ca21 (0~0~2) reflection is present, related to the $c$-axis growth of the film. The absence of any other orientation or spourius phase is confirmed by the high resolution scan (inset) performed in a shorter range around the Ca21 reflection.  (b) Rocking curve of the Ca21 (0~0~2) reflection. (c)-(d) High-resolution reciprocal space maps of the LSAO (2~0~0)  and Ca21 (0~0~2) reflections. The $c$-axis of the Ca21 is perfectly aligned to the out of plane direction of the substrate (they have the same $q_{\parallel}$).}
\label{fig:Fig1_XRD}
\end{figure}

The $\omega$-2$\theta$ scan [see Fig. \ref{fig:Fig1_XRD}(a)] confirm that the film is crystalline and $c$-axis oriented, without any hint of spurious phase, or misoriented domain. The high-resolution symmetric reciprocal space maps [see Figs. \ref{fig:Fig1_XRD}(c)-(d)] show indeed that the Ca21 (0~0~2) reflection is aligned to the LSAO (2~0~0) reflection, i.e. to the normal direction of the substrate ($q_{\parallel}^{Ca21} = q_{\parallel}^{LSAO}$). This occurrence, together with the full width at half maximum of only $\approx 0.03^{\circ}$ for the Ca21 (0~0~2) reflection [see Fig. \ref{fig:Fig1_XRD}(b)], supports the high texture of the film. Therefore the use of a substrate with small in-plane mismatch $\delta^{\mathrm{m}}$ with the film (with $\delta^{\mathrm{m}}=1-x_{\mathrm{f}}/x_{\mathrm{s}}$, and $x_{\mathrm{f}}$ and $x_{\mathrm{s}}$ in-plane lattice parameters respectively of film and substrate, see Table \ref{lattice}), prevents the distortion of the CuO$_4$ plaquettes within the Ca21 unit cell. Such distortion, causing a buckling of the atomic planes, is vice versa generally occurring in perovskite thin films, when they are grown on substrates with larger lattice mismatch \cite{gebhardt2007formation, vailionis2011misfit, weber2016multiple, arpaia2019untwinned}. Finally, the length of the $c$-axis parameter has been estimated from the position of the Ca21 (0~0~2) reflection: its value, $q^{-1}_{\perp} = 3.262$~\AA, is very close to the bulk value.

\begin{figure*}[htbp]
\includegraphics[width=1.9\columnwidth]{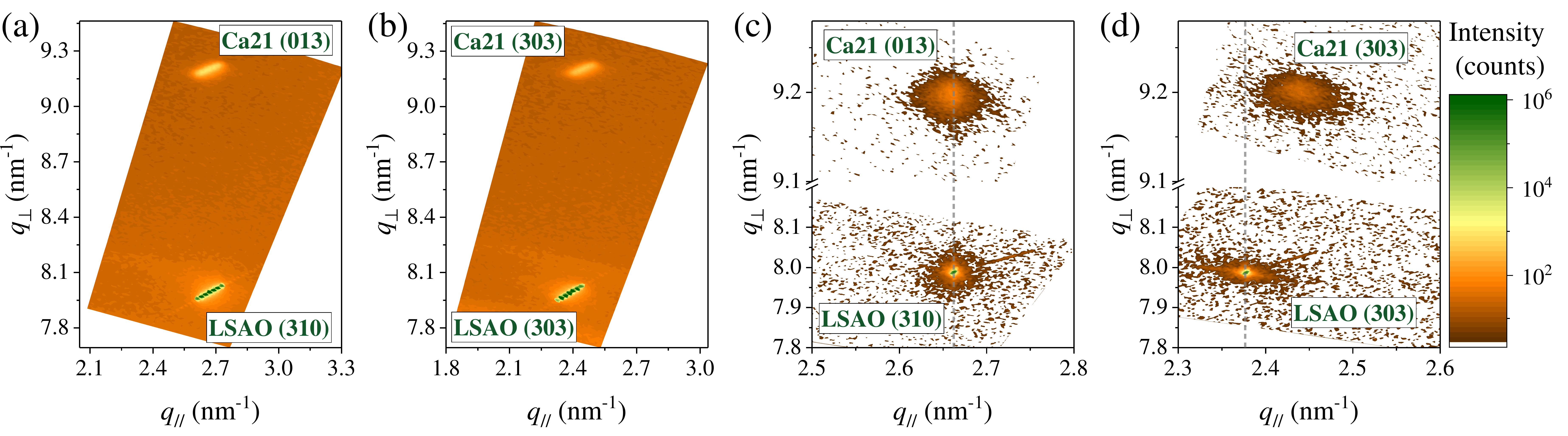}
\caption{Asymmetric XRD measurements on the Ca$_2$CuO$_3$ (30 nm)/LaSrAlO$_4$ (1~0~0). (a)-(b) Low-resolution reciprocal space maps, along the two in-plane LSAO directions. Along the [0~1~0] direction [panel (a)] only the Ca21 (0~1~3) reflection ($b$-axis) is present, while along the [0~0~1] direction [panel (b)] the Ca21 (3~0~3) is the only measured reflection ($a$-axis). Where absent, the position of the Ca21 (0~1~3) and (3~0~3) reflections has been marked in the maps by a yellow cross. The film is totally detwinned, with the following film-substrate relations: Ca21(1~0~0)$\parallel$LSAO(0~0~1), Ca21(0~1~0)$\parallel$LSAO(0~1~0), Ca21(0~0~1)$\parallel$LSAO(1~0~0). (c)-(d) High-resolution reciprocal space maps, along the two in-plane LSAO directions, used to determine the length of the in-plane Ca21 parameters and the strain state.}
\label{fig:Fig2_XRD}
\end{figure*}

To determine the twinning state of the Ca21 thin film, as well as the film-substrate orientation relations and the strain conditions, we have explored by both low- and  high-resolution reciprocal space maps the asymmetrical (0~1~3) and (3~0~3) Ca21 reflections. These two reflections, associated respectively to the $b$-axis, where the CuO$_3$ chains are oriented, and to the $a$-axis, are characterized by almost identical $q_{\perp}$, and similar $q_{\parallel}$ values (${\Delta}q_{\parallel} = 0.2$ nm$^{-1}$). The maps have been taken both along the [0~1~0] and the [0~0~1] in-plane LSAO directions. Along the [0~1~0] LSAO direction, only the Ca21 (0~1~3) reflection is present; vice versa, along the [0~0~1] LSAO direction, the Ca21 (3~0~3) reflection is dominant [see Figs. \ref{fig:Fig2_XRD}(a)-(b)].

The film is therefore totally detwinned: the random exchange of the in-plane parameters is eliminated, and the material properties, related to the presence of CuO$_3$ chains along the $b$-axis, can be singled out. The one-dimensional nature of the Cu21 is preserved in thin film form, down to thickness of few unit cells. Such film-substrate orientation relationship has severe implications on the way the strain is applied on the Cu21 film. Along the [0~1~0] LSAO $\parallel$ [0~1~0] Ca21 direction, where the mismatch is negligible ($\delta^{\mathrm{m}}\approx -0.7$\%), the Ca21 (0~1~3) reflection is aligned to the LSAO (3~1~0) reflection: the $b$-axis of the Ca21 is totally strained on the $b$-axis of the LSAO, compressing its bulk value down to $q^{-1}_{\parallel} = 3.758$~\AA [see Fig. \ref{fig:Fig2_XRD}(c)]. Along the [0~0~1] LSAO $\parallel$ [1~0~0] Ca21 direction, the mismatch is instead larger  ($\delta^{\mathrm{m}}\approx 3$\%): here, the Ca21 (3~0~3) reflection is misaligned with respect to the LSAO (3~0~3) reflection [see Fig. \ref{fig:Fig2_XRD}(d)], since the length of the Ca21 $a$-axis parameter ($q^{-1}_{\parallel} = 12.288$~\AA) approaches that of the LSAO $a$-axis, though staying much shorter than that. The tensile stain along the $a$-axis compensates the compressive one along the $b$-axis, in agreement with the negligible change of the $c$-axis parameter, observed via $\omega$-2$\theta$ scan.

The Cu21 and LSAO lattice parameters, in bulk form as well as in our films, are summarized in Table \ref{lattice}.

\begin{table}
\caption{\label{lattice} The lattice parameters of our Cu21 thin films, as determined by the XRD structural characterization, are compared with those of the bulk Cu21, and of the LSAO substrate.}
\begin{tabular}{|c|c|c|c|}
\hline \hline
material & $a$ (\AA) & $b$ (\AA)  & $c$ (\AA)\\
\hline
LaSrAlO$_4$ & 3.756 & 3.756 & 12.617\\
Ca$_2$CuO$_3$ bulk \cite{kondoh1988magnetic} & 12.262 & 3.783 & 3.263\\
Ca$_2$CuO$_3$ film & 12.288 & 3.758 & 3.262\\
\hline
\end{tabular}
\end{table}

\section{Theoretical model and methods}
\label{appendix:theo_model}
Ca$_2$CuO$_3$ is a strong charge-transfer insulator, with one hole in the $d_{x^2-y^2}$ orbital.
Hence spin and orbital excitations are described by a Kugel-Khomskii Hamiltonian with the generic form described by Eq.~\eqref{equ:kk_ham}.
The operators $A(T^\beta_i,T^\alpha_j)$ and $K(T^\beta_i,T^\alpha_j)$ describe the dynamics of the orbital degrees of freedom and are obtained by the second order perturbation theory from the multi-orbital Hubbard model, assuming: (i) dominant onsite Coluomb repulsion (`Hubbard' $U$ is much larger than the hopping between the ground state orbital $t_1$), i.e. omitting charge fluctuations; (ii) strong crystal field splitting, polarizing the ground state in the lower orbital;  and (iii) one excited orbital at most.
Altogether, the operators read
\begin{widetext}
\begin{align} \label{equ:supp_kk_ham_A}
   A(T^\beta_i,T^\alpha_j) = \frac{2t_1^2}{U} \frac{1}{1-\left( \frac{J_H}{U} \right)^2} \left( T^z_i -\frac{1}{2} \right) \left( T^z_j -\frac{1}{2} \right)
   +\frac{2t_1 t_2}{U^\prime} \frac{1}{1-\left( \frac{J_H}{U^\prime} \right)^2} \left[ T_i^+ T_j^- + \text{h.c.} \right]\nonumber\\
   + \frac{ ( t_1^2 + t_2^2 )}{U^\prime} \frac{\frac{J_H}{U^\prime}}{1-\left( \frac{J_H}{U^\prime} \right)^2} \left[ \left( T^z_i -\frac{1}{2} \right) \left( T^z_j +\frac{1}{2} \right)+ (i\leftrightarrow j)  \right]\; ,
\end{align}
and
\begin{align} \label{equ:supp_kk_ham_K}
   K(T^\beta_i,T^\alpha_j) =
    -\frac{2t_1^2}{U} \frac{1}{1-\left( \frac{J_H}{U} \right)^2} \left( T^z_i -\frac{1}{2} \right) \left( T^z_j -\frac{1}{2} \right)
    +\frac{2t_1 t_2}{U^\prime} \frac{ \frac{J_H}{U^\prime} }{1-\left( \frac{J_H}{U^\prime} \right)^2} \left[ T_i^+ T_j^- + \text{h.c.} \right]\nonumber \\
 + \frac{t_1^2 + t_2^2}{U^\prime} \frac{ 1 }{1-\left( \frac{J_H}{U^\prime} \right)^2} \left[ \left( T^z_i -\frac{1}{2} \right) \left( T^z_j +\frac{1}{2} \right)+ (i\leftrightarrow j)  \right].
\end{align}
\end{widetext}
Here, $t_1$ and $t_2$ are the hopping parameters of the ground state, i.e., between neighboring $d_{x^2-y^2}$ orbitals, and the excited orbital, i.e. $d_{xy}$, $d_{yz}$ or $d_{xz}$, respectively. ($\vec{S}_i$, $T^\alpha_i$ and all other model parameters are defined in the main text.)
As an approximation we set the inter-orbital repulsion $U^\prime = U - 2J_H$, as in the case of the atomic orbitals subject
to a spherically symmetric potential \cite{griffith_theory_1971}.
Note however, that a small departure from this relation does
not alter the results significantly~\cite{heverhagen_spinon_orbiton_2018}.

To calculate $dd$-excitation spectrum probed by RIXS we define the spectral function for orbital excitations and spin-orbital excitations.
The orbital spectral function is
\begin{align} \label{equ:supp_sqw}
   O(q,\omega) = \frac{1}{\pi} \Im \bra{\text{gs}} T^x_{-k} \frac{1}{\omega + E_\text{gs} - H - i\Gamma_L} T^x_k \ket{\text{gs}},
\end{align}
where $T^x_k$ is the Fourier transform of $T^x_i$, $|\text{gs} \rangle$ is the ground state of Eq. \eqref{equ:kk_ham}, and $\Gamma_L$ is the Lorentzian broadening to mitigate finite-size artefacts.
We set $\Gamma_L= 0.06\, \text{eV}$. The spin-orbital excitation function $SO(q,\omega)$ is defined analogously, with $T^x_i$ replaced by $S^z_iT^x_i$.

We calculate the excitation spectra using exact diagonalization in combination with the spin cluster perturbation theory \cite{ovchinnikov_cluster_2010} on a 20 site chain. Additionally, we derive in section \ref{appendix:eff_ham} an effective Hamiltonian, which yields an analytic relation between bandwidth of the excitation and the Hund's coupling.

\begin{figure*}[htbp]
   \includegraphics{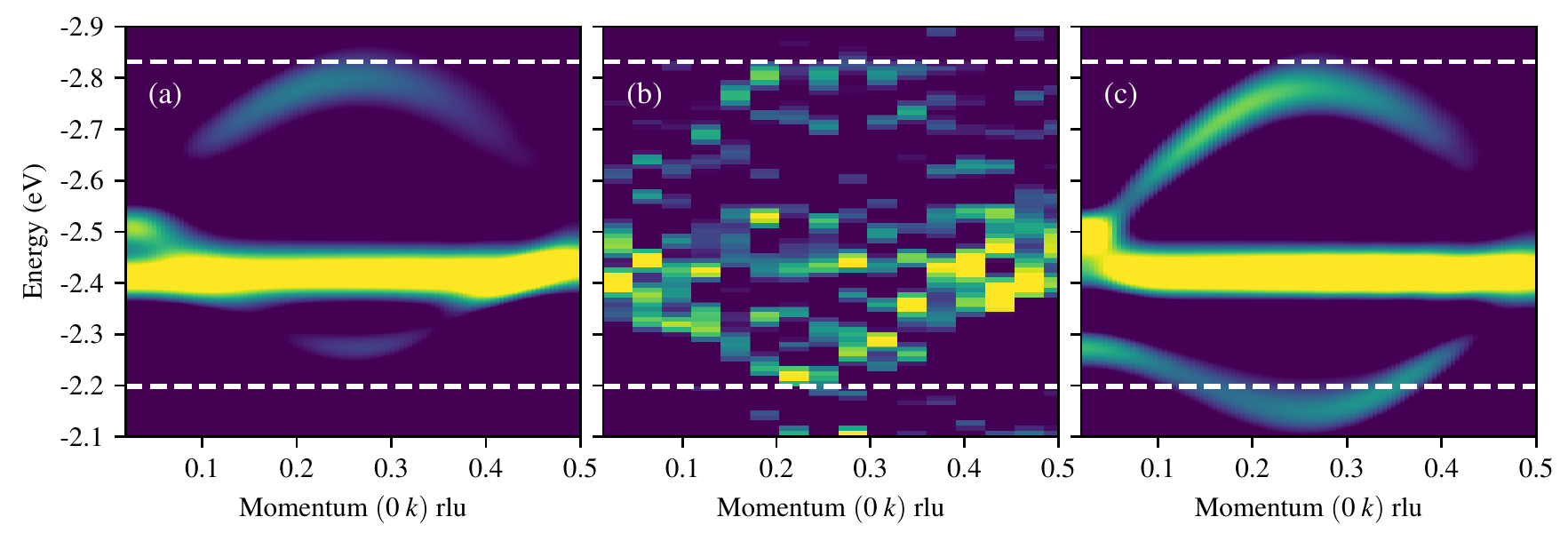}
   \caption{Theoretical (a,c) and experimental (b) second derivatives RIXS intensity maps for $\pi$ incoming polarization to estimate the error of $J_H$ for $d_{yz}$ orbital. Panel (a) and (c) show minimal $J_H=1.0\,t_1$ and maximal $J_H=1.4\,t_1$ for $d_{yz}$, respectively. Other parameters are kept constant and can be found in Table\,\ref{table:Parameters}. White lines are guides to the eye. }
\label{fig:error_estimates}
\end{figure*}

The `best fit' microscopic parameters for the Kugel-Khomskii Hamiltonian as well as the corresponding error range listed in Table\,\ref{table:Parameters} are obtained by direct comparison of the second derivative RIXS intensity maps.
This is illustrated in Fig.~\ref{fig:error_estimates} for the case of $J_H$ of $d_{yz}$. Panel (a) and (c) show the theoretical second derivative RIXS intensity maps for the lower and upper bound of the error range, respectively.
Panel (b) shows the experimental one for comparison.
Although the shape for minimal (maximal) $J_H$ matches the experimental one, the bandwidth is clearly too small (large).
Error ranges for other parameters and orbitals are estimated similarly.

\section{Effective Hamiltonian for spin-orbital excitation}
\label{appendix:eff_ham}

Here we derive an effective Hamiltonian, which captures accurately the spin-orbital excitation spectrum $SO(q,\omega)$ of Hamiltonian \eqref{equ:kk_ham} and which can be mapped onto the $t$--$J$ model.
The latter enables us to find an analytic relation between the Hund's coupling and the bandwidth $W$ of the excitation spectrum.

\begin{figure*}[htbp]
   \includegraphics{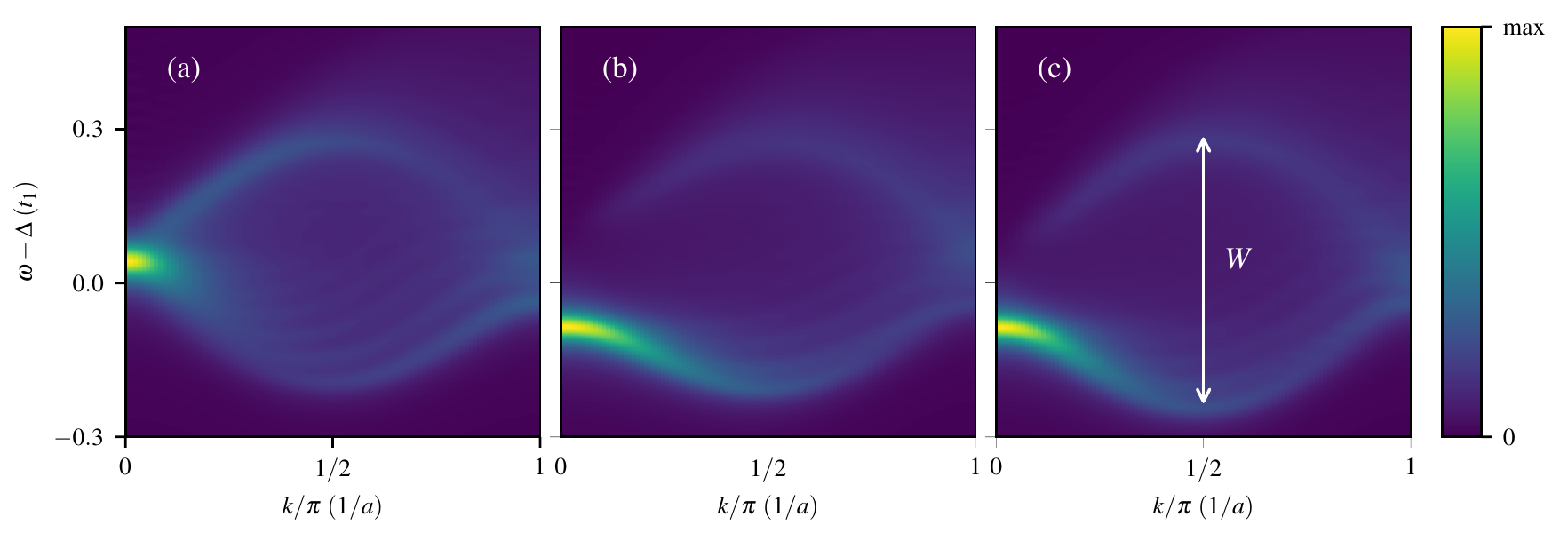}
   \caption{Orbital (a) and spin-orbital (b) spectral function of the full Hamiltonian Eq.~(\ref{equ:kk_ham}) with the orbital operators given by Eqs.~(\ref{equ:supp_kk_ham_A}-\ref{equ:supp_kk_ham_K}); (c) spin-orbital spectral function of the effective Hamiltonian, i.e. the full Hamiltonian Eq.~(\ref{equ:kk_ham}) but with the orbital operators replaced by Eqs.~(\ref{equ:eff_kk_ham_A}- \ref{equ:eff_kk_ham_K}). $W$ indicates the bandwidth of the excitation. Parameters for the $d_{yz}$ orbital excitation in Ca21 are used.
} \label{fig:supp_spectra}
\end{figure*}

The ground state of Hamiltonian \eqref{equ:kk_ham}  is antiferromagnetically (AF) and ferro-orbitally (FO) ordered.
Hence we approximate the ground state $\Ket{\text{gs}}$ at a bond $\braket{i,j}$ as spin-singlet with an FO order
\begin{equation} \label{equ:twositeGS}
   \Ket{\text{gs}} \propto \Ket{\tikzstate{lowup}{lowdown}} - \Ket{\tikzstate{lowdown}{lowup}}.
\end{equation}
The excited state for $SO(q,\omega)$ (see Fig.~\ref{fig:supp_spectra}) excitation is then given by
\begin{equation} \label{equ:excspinorb}
   S^z_jT^+_j \Ket{\text{gs}} \propto \Ket{\tikzstate{lowup}{highdown}} + \Ket{\tikzstate{lowdown}{highup}}.
\end{equation}
Thus, we are now allowed to rewrite the parts in Hamiltonian \eqref{equ:kk_ham} which hinder the mapping onto a $t$--$J$ model.
This yields an effective Hamiltonian of the form \eqref{equ:kk_ham}, with the modified orbital operators
\begin{widetext}
\begin{align} \label{equ:eff_kk_ham_A}
   \widetilde A(T^\beta_i,T^\alpha_j) = \frac{2t_1^2}{U} \frac{1}{1- \left(\frac{J_H}{U}\right)^2 } \left( T^z_i -\frac{1}{2} \right) \left( T^z_j -\frac{1}{2} \right)
   +\frac{2t_1 t_2}{U^\prime} \frac{1}{1- \frac{J_H}{U^\prime} } \left[ T_i^+ T_j^- + \text{h.c.} \right],
\end{align}
and
\begin{align} \label{equ:eff_kk_ham_K}
   \widetilde K(T^\beta_i,T^\alpha_j) =
    -\frac{2t_1^2}{U} \frac{1}{1-\left( \frac{J_H}{U} \right)^2} \left( T^z_i -\frac{1}{2} \right) \left( T^z_j -\frac{1}{2} \right)
 + \frac{t_1^2+t_2^2}{U^\prime} \frac{ 1 }{1- \frac{J_H}{U^\prime} } \left[ \left( T^z_i -\frac{1}{2} \right) \left( T^z_j +\frac{1}{2} \right)+ (i\leftrightarrow j)  \right].
\end{align}
\end{widetext}
Note that the effective Hamiltonian can now be mapped onto a $t$--$J$ model~\cite{Wohlfeld2011}.

The mapping on the $t$--$J$ model gives analytic expressions for upper and lower boundaries of the spinon-orbiton continuum \cite{Brunner2000,Suzuura1997}, which read
\begin{subequations}
   \begin{align}
      \epsilon_{\text{lo}}(k) =& \Delta + \mathcal{J} - 2 \frac{t_1^2+t_2^2}{U^\prime}  \frac{1}{(1-J_H/U^\prime) } - \nonumber \\
      &\begin{cases}
      \sqrt{ \mathcal{J}^2 + 4 \tilde{t}^2 - 4\tilde{t}\mathcal{J} \cos(k)}&  ,k<k_0\\
      2\tilde{t} \sin(k) &, k>k_0
\end{cases}, \label{equ:tj_model_compact_support_lo}\\
      \epsilon_{\text{up}}(k) =& \Delta + \mathcal{J} - 2 \frac{t_1^2+t_2^2}{U^\prime}  \frac{1}{(1-J_H/U^\prime) }  + \nonumber \\
      &\begin{cases}
      \sqrt{ \mathcal{J}^2 + 4 \tilde{t}^2 - 4\tilde{t}\mathcal{J} \cos(k)}&  ,k>k_0\\
      2\tilde{t} \sin(k) &, k<k_0
\end{cases}, \label{equ:tj_model_compact_support_up}
\end{align}
\end{subequations}
Here $k_0$ is given by {$\cos(k_0)=\mathcal{J}/(2\tilde{t})$} and the rescaled hopping  and rescaled superexchange constants are given by
\begin{equation} \label{equ:tandJ}
   \tilde{t}= 2 \frac{t_1  t_2}{U^\prime}  \frac{1}{(1- J_H/U^\prime)} \quad \text{and} \quad \mathcal{J}=  J_\text{AF}  \frac{1}{1-(J_H/U)^2}
\end{equation}
respectively.
The bandwidth is then given by
\begin{equation} \label{equ:bandwidth}
   W = \epsilon_\text{up}(k=\pi/2) - \epsilon_\text{lo}(k=\pi/2) = 4 \tilde{t}.
\end{equation}
This relation can be used to extract Hund's coupling from the bandwidth of the excitation spectrum.

Fig.~\ref{fig:supp_spectra} shows the orbital (a) and the spin-orbital (b) excitation spectrum of the full Hamiltonian as well as the spin-orbital (c) excitation spectrum of the effective Hamiltonian, for parameters modelling the $d_{yz}$ orbital excitation in Ca21.
The effective Hamiltonian reproduces the shape as well as the intensity of the spin-orbital excitations of the full Hamiltonian.
RIXS measures a combination of orbital and spin-orbital excitation, however due to the same bandwidth of $O(k,\omega)$ and $SO(k,\omega)$ this does not affect the expression of the bandwidth~\eqref{equ:bandwidth}.

\bibliography{Biblio_CCO_1D_RIXS}

\end{document}